\documentclass[journal=jcisd8,manuscript=article]{achemso}

\usepackage[version=3]{mhchem} 
\usepackage{kotex}
\usepackage{color}
\usepackage{amsmath,amssymb}
\DeclareMathOperator{\E}{\mathbb{E}}

\newcommand{\norm}[1]{\left\lVert#1\right\rVert}
\usepackage{amsfonts} 

\author{Seung Hwan Hong}
\affiliation[KAIST]{Department of Chemistry, KAIST, 291 Daehak-ro, Yuseong-gu, Daejeon 34141, Republic of Korea}
\author{Jaechang Lim}
\affiliation[KAIST]{Department of Chemistry, KAIST, 291 Daehak-ro, Yuseong-gu, Daejeon 34141, Republic of Korea}
\author{Seongok Ryu}
\affiliation[KAIST]{Department of Chemistry, KAIST, 291 Daehak-ro, Yuseong-gu, Daejeon 34141, Republic of Korea}
\author{Woo Youn Kim}
\affiliation[KAIST]{Department of Chemistry, KAIST, 291 Daehak-ro, Yuseong-gu, Daejeon 34141, Republic of Korea}
\alsoaffiliation[KI4AI]{KI for Artificial Intelligence, KAIST, 291 Daehak-ro, Yuseong-gu, Daejeon 34141, Republic of Korea}
\email{wooyoun@kaist.ac.kr}

\title[An \textsf{achemso} demo]
  {Molecular Generative Model Based On Adversarially Regularized Autoencoder
  }

\abbreviations{IR,NMR,UV}

\begin{document}
\begin{tocentry}

Some journals require a graphical entry for the Table of Contents.
This should be laid out ``print ready'' so that the sizing of the
text is correct.

Inside the \texttt{tocentry} environment, the font used is Helvetica
8\,pt, as required by \emph{Journal of the American Chemical
Society}.

The surrounding frame is 9\,cm by 3.5\,cm, which is the maximum
permitted for  \emph{Journal of the American Chemical Society}
graphical table of content entries. The box will not resize if the
content is too big: instead it will overflow the edge of the box.

This box and the associated title will always be printed on a
separate page at the end of the document.

\end{tocentry}

\begin{abstract}
Deep generative models are attracting great attention as a new promising approach for molecular design.
All models reported so far are based on either variational autoencoder (VAE) or generative adversarial network (GAN). 
Here we propose a new type model based on an adversarially regularized autoencoder (ARAE). 
It basically uses latent variables like VAE, but the distribution of the latent variables is obtained by adversarial training like in GAN. 
The latter is intended to avoid both inappropriate approximation of posterior distribution in VAE and difficulty in handling discrete variables in GAN. Our benchmark study showed that ARAE indeed outperformed conventional models in terms of validity, uniqueness, and novelty per generated molecule. We also demonstrated successful conditional generation of drug-like molecules with ARAE for both cases of single and multiple properties control. As a potential real-world application, we could generate EGFR inhibitors sharing the scaffolds of known active molecules while satisfying drug-like conditions simultaneously.      

\end{abstract}

\section{Introduction}

One of the prime goals of chemistry is to make novel molecules with desired properties for various purposes. 
Inspired by the great success of deep generative models in computer vision tasks,\cite{kingma2013auto, goodfellow2014generative} molecular generative models have emerged as a new promising approach for efficient molecular design. \cite{Chen2018,Sanchez-Lengeling2018} 
Diverse models have been proposed for materials design and de novo drug design with promising results.\cite{gomez2018automatic, segler2017generating, popova2018deep, lim2018molecular, li2018multi, you2018graph, de2018molgan, gupta2018,bjerrum2018improving,Ikebata2017,kang2019,polykovskiy2018,guimaraes2017objective,olivecrona2017,neil2018exploring} 
Their key idea is to estimate the distribution of molecules for a specific purpose and then to sample unseen molecules with target properties from the estimated distribution. 
So far, generative adversarial networks (GANs)\cite{goodfellow2014generative} and latent variable models have been widely used for that purpose. 
GANs directly estimate the distribution of true input data by adversarial training of a generator and a discriminator. 
On the other hand, latent variable models such as variational autoencoders (VAEs)\cite{kingma2013auto} estimate the distribution of latent variables corresponding to input data and generate new molecules by decoding latent variables sampled. 
Furthermore, one can incorporate desired properties directly in the generation process by estimating the conditional distribution of molecules involving specific target properties. 

For instance, ChemicalVAE\cite{gomez2018automatic} is one of the latent variable models. Its latent space is jointly trained with a deep neural network for prediction of a target property. 
As a result, molecules with desired properties can be designed by optimizing the target property on the latent space. 
The feasibility of molecular generative model demonstrated by ChemicalVAE triggered the active development of various models with similar concepts. \cite{kusner2017grammar, lim2018molecular, kang2018conditional} 
However, VAE-based models often produce unnatural molecules, leading to the low validity of generated molecules. \cite{de2018molgan} 
That might be because a unknown posterior distribution is approximated by a given family of surrogate distributions, but inappropriate approximation can cause latent variables being decoded to unnatural molecules. 
Since GANs directly estimate the distribution of true input data via adversarial training, they can avoid limitations arisen from such an approximation. ORGAN\cite{guimaraes2017objective} and MolGAN\cite{de2018molgan} are prototypes of GAN-based molecular generative models. 
Indeed, GAN-based models show much improved validity. 
In contrast to computer vision tasks, however, molecular structures are expressed by the discrete categorical representation of atomic symbols. 
GAN-based models are troublesome to deal with such a discrete variable because of difficulty in estimating their distribution. 
As a result, molecules generated by ORGAN and MolGAN were not diverse, leading to low uniqueness.

Having considered the above facts, it seems desirable to selectively take each advantage of GAN- and VAE-based models to accomplish both high validity and uniquness.    
For this purpose, we propose to use the platform of adversarially regularized autoencoder (ARAE)\cite{junbo2017adversarially} for molecular design.
This model is grounded on the spirit of latent variable models that transform the discretized input of molecular structures to continuous latent representations.
The key distinct feature of this model is to adopt the adversarial training used in GANs to estimate the distribution of latent variables, which makes it avoid the problem caused by the posterior approximation. As a result, one can achieve a high valid rate as well as high uniqueness in molecular generation. 
For conditional generation, we disentangle the information of molecular properties from latent vectors. Then, the target molecular properties are injected independently with latent vectors into the decoder. 
Thus, we could produce unseen molecules having the designated molecular properties with a high success rate. 

We demonstrate the usefulness of our model with the following examples:
\begin{itemize}
    \item We verify the high performance of our model in estimating a latent vector distribution by showing the validity, uniqueness, and novelty of generated molecules. We also test smoothness of latent space by interpolating between two vectors in the latent space.
    \item We show the feasibility of the simultaneous control of multiple properties with a high success rate. 
    \item As a possible practical application, we demonstrate that our model can be used for \textit{de novo} design of hit compounds in drug discovery with the example of epidermal growth factor receptor (EGFR) inhibitors. 
\end{itemize}

\section{Previous works}
To compare ARAE with GAN and VAE and its technical advantages for molecular applications, we briefly introduce about each method. Then, we describe our implementation of the ARAE modified for molecular generations in detail.
\subsection{Generative adversarial network}
GANs \cite{goodfellow2014generative} estimate the distribution of input samples, $p_{r}(\textbf{x})$, through the adversarial training of a generator network and a discriminator network. 
By taking a random variable $\textbf{s}$ drawn from a prior distribution $p(\textbf{s})$, the generator $g_{\psi}$ parameterized by $\psi$ produces new samples $g_{\psi}(\textbf{s})$ of the distribution $p_g(\textbf{x})$. 
To estimate the distribution of true data by generating samples with the generator, Wassestein GAN is used to minimize the gap between the true and generated distributions ($p_{r}(\textbf{x})$ and $p_{g}(\textbf{x})$, respectively).\cite{arjovsky2017wasserstein} 
Thus, the training objective is given by
\begin{equation} \label{eqn:eq1}
    \min_{\psi} \text{W}(p_{r}, p_{g}) = \min_{\psi} \max_{w} \E_{\textbf{x} \sim p_{r}(\textbf{x})} [f_{w}(\textbf{x})] - \E_{ \textbf{s} \sim p(\textbf{s}) } [f_{w}(g_{\psi}(\textbf{s}))],
\end{equation}
where $f_w$ is a discriminator (critic) function parameterized by $w$ satisfying the 1-Lipschtiz continuity $\norm{f_w} \leq 1$. 
As a result of the adversarial training of these two networks, the distributions of data samples $p_{r}(\textbf{x})$ and of generated samples $p_g(\textbf{x})$ become equivalent, i.e., $p_{r}(\textbf{x}) =  p_g(\textbf{x})$. 

Learning discretized representations such as molecular structures with GANs often fails because outputs can be easily degenerated into training data. \cite{kusner2016gans, junbo2017adversarially} 
This drawback causes low efficiency for the generation of unseen molecules with GAN-based models. 

\subsection{Variational autoencoder}
Instead of directly modeling the distribution of input data, latent variable models infer the distribution of latent variables (the posterior). 
However, it is intractable to find an exact posterior distribution in most cases. 
VAEs approximate the posterior distribution with the variational distribution $q_{\theta}(\textbf{z}|\textbf{x})$ which is an output of the encoder parameterized with $\theta$, given an input $\textbf{x}$, 
and the decoder parameterized by $\phi$ reconstructs the inputs from the latent variables drawn from the posterior.\cite{kingma2013auto} 
The minimization objective of VAEs is given by
\begin{equation} \label{eqn:eq2}
    \min_{\theta, \phi} \mathbb{E}_{\textbf{x} \sim p_r(\textbf{x})} [\mathcal{L}_{rec}(\theta, \phi) + \text{KL}(q_{\theta}(\textbf{z}|\textbf{x}) || p(\textbf{z}))],
\end{equation}
where $\mathcal{L}_{rec}(\theta, \phi) = \mathbb{E}_{q_{\theta}(\textbf{z}|\textbf{x})}[- \log{p_{\phi}(\textbf{x}|\textbf{z})}]$ is the reconstruction loss and 
$\text{KL}(q_{\theta}(\textbf{z}|\textbf{x}) || p(\textbf{z}) )$ is the Kullback-Leiber (KL) divergence between the variational distribution $q_{\theta}(\textbf{z}|\textbf{x})$ and the prior distribution $p(\textbf{z})$. 
Minimizing the second term makes the two distributions as similar as possible. 
Hence, the posterior distribution can be approximated by imposing it on the prior distribution. 

Since VAEs approximate the posterior distribution with a predefined prior (a surrogate family of distributions), they can readily estimate the distribution of latent variables.  
However, the drawback of using the VAEs is well-known; a latent space can have holes in which latent vectors are not matched to true data points. \cite{makhzani2015adversarial}
That causes the generation of chemically unrealistic molecules. \cite{gomez2018automatic}
The main reasons of the hole existing problem are as follows.
First, the true posterior distribution may not be well approximated by a given prior, such as a normal distribution.
Second, minimizing the KL-divergence between two distributions is not suitable if the posterior distribution is multi-modal. \cite{murphy2012machine, goodfellow2016nips}
In such cases, using VAEs may not be a good approach to model a latent variable distribution. 

\subsection{Adversarially regularized autoencoder}
The ARAE method was proposed to address the aforementioned problems of the GANs and the VAEs. \cite{junbo2017adversarially}
It is basically a latent variable model which adopts an encode-decoder architecture, but the posterior distribution is estimated by adversarial training. 
The encoder network parameterized by $\theta$ outputs the distribution of true latent variables $\textbf{z} = \textrm{enc}_{\theta}(\textbf{x}) \sim p_{\theta} (\textbf{z})$ from the given inputs. The decoder network parameterized by $\phi$ reconstructs the input from the latent variable drawn from the posterior.
By following the idea of GANs, the generator parmeterized by $\psi$ outputs the distribution of generated random variables $\tilde{\textbf{z}} = g_{\psi}(\textbf{s}) \sim p_{\psi}(\tilde{\textbf{z}})$, where $\textbf{s}$ is a random variable drawn from the prior distribution $p(\textbf{s})$.  
Since ARAE aims to estimate the posterior distribution by generating a similar distribution with that of the generator, 
the training objective is given by 
\begin{equation} \label{eqn:eq3}
    \min_{\theta, \phi, \psi} \mathbb{E}_{\textbf{x} \sim p_{r}(\textbf{x})} [\mathcal{L}_{rec}(\theta, \phi) + \text{W}(p_{\theta} (\textbf{z}), p_{\psi}(\tilde{\textbf{z}}))],
\end{equation}
where the reconstruction loss is written as
\begin{equation} \label{eqn:eq4}
    \mathcal{L}_{rec}(\theta, \phi) = \mathbb{E}_{\textbf{z} \sim p_{\theta} (\textbf{z})}[- \log{p_{\phi}(\textbf{x}|\textbf{z})}]
\end{equation}
and the Wasserstein-1 distance between the two distributions is written as
\begin{equation} \label{eqn:eq5}
    \text{W}(p_{\theta} (\textbf{z}), p_{\psi}(\tilde{\textbf{z}})) = \max_{w} \mathbb{E}_{\textbf{z} \sim p_{\theta} (\textbf{z})} [f_{w}(\textbf{z})] - \mathbb{E}_{\tilde{\textbf{z}}  \sim p_{\psi}(\tilde{\textbf{z}})} [f_{w}(\tilde{\textbf{z}})],
\end{equation}
with the 1-Lipschtiz continuity $\norm{f_w} \leq 1$.
As a result of training, the two distributions $p_{\theta}$ and $p_{\psi}$ become identical, and we can generate new samples by using random variables sampled from  $p_{\psi}$ as an input to the decoder. 

\section{Methods}

\begin{figure}[] 
    \includegraphics[width=0.49\textwidth,trim={0cm 0 0cm 0},clip]{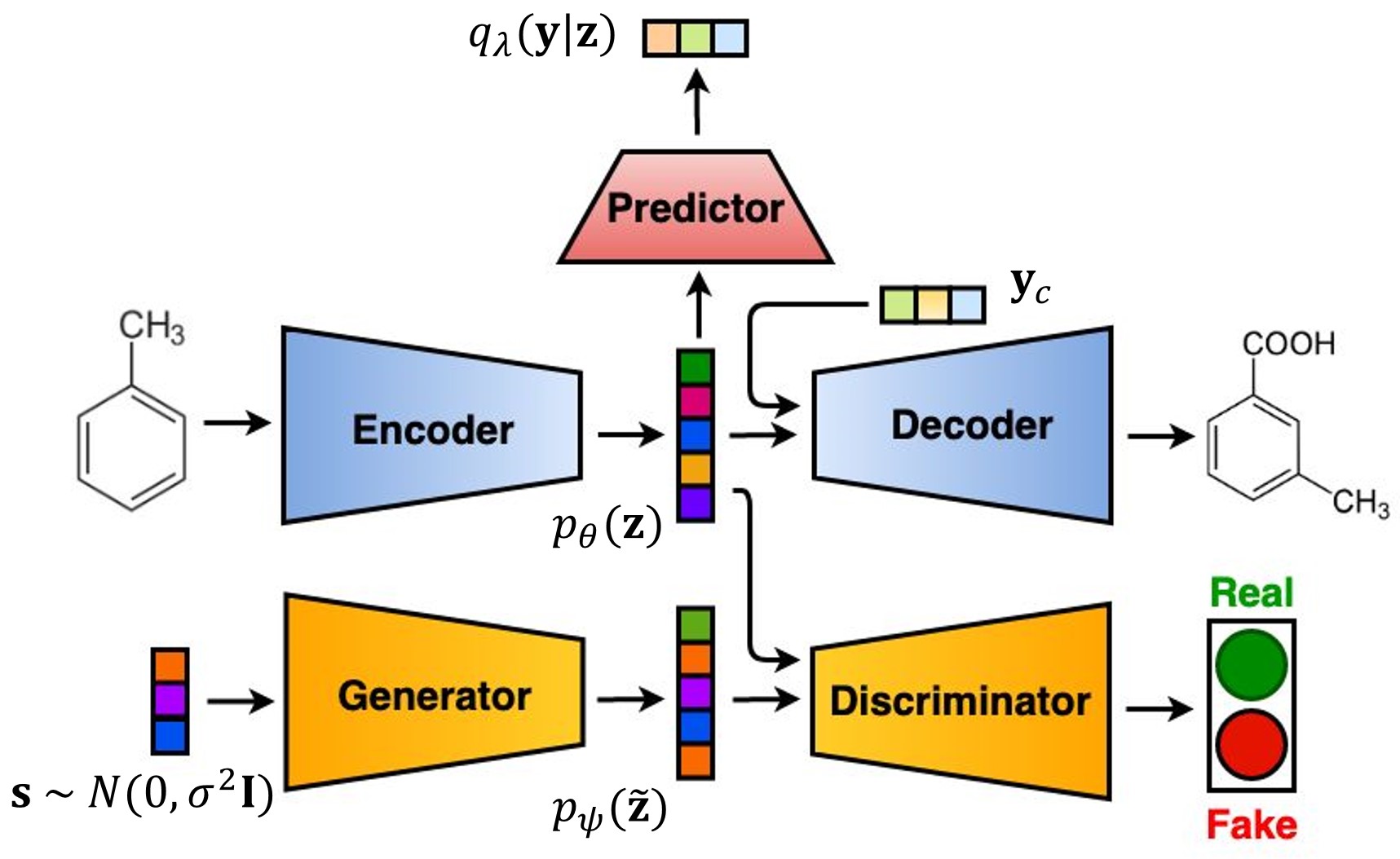}
    \caption{The architecture of CARAE for molecular generations. 
    The encoder embeds the SMILES representation of molecular structure $\textbf{x}$ to the latent vector $\textbf{z}$, 
    and the decoder reconstructs the molecular structures from the latent vector. 
    As a result of the adversarial training, two distributions $p_{\theta}(\textbf{z})$ and $p_{\psi}(\tilde{\textbf{z}})$ become equivalent. 
    The predictor is trained to predict an original molecular property $\textbf{y}$ and separate this information from the latent vector by minimizing the mutual information term in eq. \eqref{eqn:eq7}. 
    In decoding phase, the specified property information $\textbf{y}_c$ is incorporated together with the latent vector to generate the molecules with specific desired property.}
    \label{fig:ARAE-architecture}
\end{figure}

We devised a new type of molecular generative model by adopting the architecture of ARAE. 
The SMILES representation of molecular structure is adopted as a discrete random variable input to the model. The raw input is first transformed to the latent representation, and its distribution is estimated by adversarial training. 
As we discussed in the previous section, this approach has the following advantages. First, it can avoid the hole-generating problem appeared in VAEs by estimating the posterior through adversarial training. As a result, the valid rate of generated molecules is expected to be high. Second, it improves a low learning ability of GANs for discrete representation like SMILES by using continuous latent variables in the adversarial training. 

In addition, we introduced an efficient conditional generation scheme of molecules. 
The key idea of conditional generative model is to estimate the distribution of data samples as latent vectors and conditions are jointly given: $p(\textbf{x}|\textbf{z},\textbf{y}_c)$. 
However, the latent vectors, which are supposed to correspond to molecular structures, would not be independent from target molecular properties.  
To control the target properties and structures in parallel, therefore, we design a model so as to disentangle the property information from the latent vectors.   
To do so, we jointly minimize the mutual information $\textrm{MI}(\textbf{z}, \textbf{y}; \theta, \lambda)$ which means the amount of the information of a target property $\textbf{y}$ obtained from the latent vector $\textbf{z}$.
Since an exact value of the mutual information is not known, 
we compute the variational mutual information\cite{agakov2004algorithm, chen2016infogan} given by 
\begin{equation} \label{eqn:eq6}
\begin{split}
    \textrm{MI}(\textbf{z}, \textbf{y}; \theta, \lambda) 
    & = \textrm{H}(\textbf{y}) - \textrm{H}(\textbf{y}|\textbf{z}) \\
    & = \mathbb{E}_{\textbf{z} \sim p_{\theta}(\textbf{z})} [\mathbb{E}_{\textbf{y} \sim p(\textbf{y}|\textbf{z})} \log p(\textbf{y}|\textbf{z})] 
    + \textrm{H}(\textbf{y}) \\
    & = \mathbb{E}_{\textbf{z} \sim p_{\theta}(\textbf{z})} [\textrm{KL}(p(\textbf{y}|\textbf{z})||q_{\lambda}(\textbf{y}|\textbf{z})) 
    + \mathbb{E}_{\textbf{y} \sim p(\textbf{y}|\textbf{z})} [\log q_{\lambda}(\textbf{y}|\textbf{z})]] + \textrm{H}(\textbf{y}) \\
    & \geq \mathbb{E}_{\textbf{z} \sim p_{\theta}(\textbf{z})} [\mathbb{E}_{\textbf{y} \sim p(\textbf{y}|\textbf{z})} [\log q_{\lambda}(\textbf{y}|\textbf{z})]]
    + \textrm{H}(\textbf{y}),
\end{split}    
\end{equation}
where $\textrm{H}(\textbf{y})$ denotes the marginal entropy, $\textrm{H}(\textbf{y}|\textbf{z})$ is the joint entropy of $\textbf{y}$ and $\textbf{z}$, and $q_{\lambda}(\textbf{y}|\textbf{z})$ is the auxiliary conditional distribution of $\textbf{y}$ for the given $\textbf{z}$, which is estimated by the predictor parameterized by $\lambda$.  
As done in the previous work\cite{chen2016infogan}, we also consider $\textrm{H}(\textbf{y})$ as a constant. 
Therefore, we set the training objective as follows:
\begin{equation} \label{eqn:eq7}
    \min_{\theta, \phi, \psi} \mathbb{E}_{\textbf{x} \sim p_r(\textbf{x})} [\mathcal{L}_{rec}(\theta, \phi) + \text{W}(p_{\theta}(\textbf{z}),
    p_{\psi}(\tilde{\textbf{z}})) + \text{VMI}(\textbf{y}, \textbf{z} ; \theta, \lambda)],
\end{equation}
where $\text{VMI}(\textbf{y}, \textbf{z} ; \theta, \lambda) = \max_{\lambda} \mathbb{E}_{\textbf{z} \sim p_{\theta}(\textbf{z})} [\mathbb{E}_{\textbf{y} \sim p(\textbf{y}|\textbf{z})} [\log q_{\lambda}(\textbf{y}|\textbf{z})]]$. 
We note that minimizing the third term makes the property information separated from the latent vector. 
In the decoding phase, the target property information $\textbf{y}_c$, which acts as a condition vector for the reconstruction of molecules, is given together with the latent vector of query molecule $\textbf{z}$. 
As a result, molecules with designated properties can be generated with independent structural control. 
We term this model as `conditional ARAE (CARAE)' hereafter. 

Figure \ref{fig:ARAE-architecture} illustrates the architecture of the CARAE model for molecular generations. 
SMILES seqeunces are transformed by the encoder into latent variables. The generator produces new samples by taking random variables from a distribution $\mathcal{N}(0, \sigma^2 \textbf{I})$. 
Then, the distributions of those two variables become as similar as possible by minimizing the first and the second term of eq. \eqref{eqn:eq7} with gradient descent optimization. 
For the CARAE model, we add the predictor network which is used to estimate the variational mutual information term, the third term in eq. \eqref{eqn:eq7}. 
In the training phase, the decoder reconstructs input molecular structures from the latent vector and property information of input molecules.
In the test phase, we can sample new molecules by tuning the latent vector which is drawn from $p_{\psi}(\tilde{\textbf{z}})$ and by specifying the desired property $\textbf{y}_c$.

\section{Results and discussion}
To train and test our model, we used the QM9 and ZINC datasets. 
The QM9 set contains 133,885 small organic molecules with up to nine heavy atoms.\cite{ramakrishnan2014quantum}
The ZINC set used in this work is same with that used in training ChemicalVAE\cite{gomez2018automatic}, which consists of 249,455 molecules randomly selected from the drug-like subset of the ZINC database\cite{irwin2012zinc}.

Before explaining experimental results, we introduce the metrics used to evaluate the performance of our model as follows:
\begin{itemize}
    \item \textbf{Validity}: the ratio of the number of valid molecules to the number of generated samples. The validity was checked by using RDKit.\cite{landrum2006rdkit} 
    \item \textbf{Uniqueness}: the ratio of the number of unrepeated molecules to the number of valid molecules.
    \item \textbf{Novelty}: the ratio of the number of molecules which are not included in the training set to the number of unique molecules.
    \item \textbf{Novel/Sample}: the ratio of the number of valid, unique, and novel molecules to the total number of generated samples.
\end{itemize}

\subsection{Performance of ARAE on molecular generation}
\begin{figure}[htb]
    \includegraphics[width=0.48\textwidth,trim={0cm 0 0cm 0},clip]{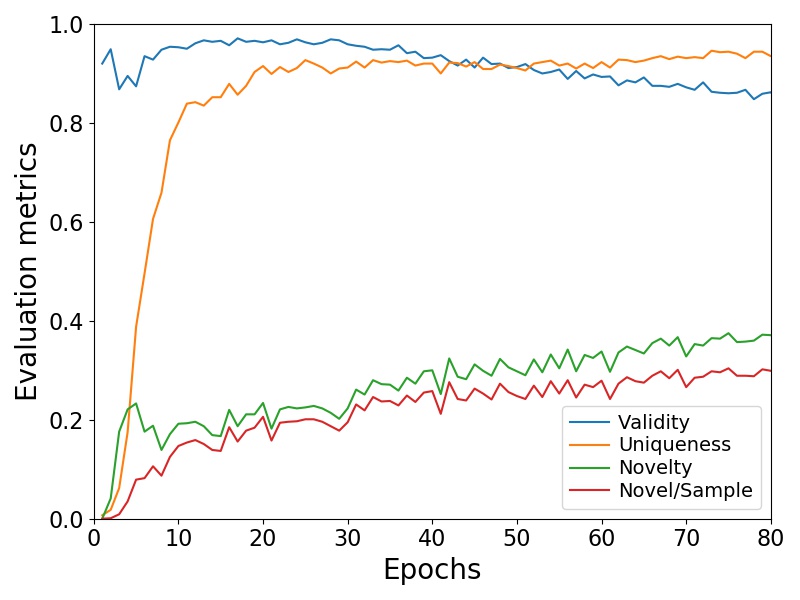}
    \caption{Convergence of the four evaluation metrics for the ARAE model trained with the QM9 dataset.}
    \label{fig:metrics}
\end{figure}

\begin{table*}[htb]
    \begin{center}
         \begin{tabular}{ccccc}
         \hline
         Method & Validity (A) & Uniqueness (B) & Novelty (C) & Novel/Sample (A$\times$B$\times$C) \\
         \hline
         ChemicalVAE   & 0.103 & 0.675 & 0.900 & 0.063 \\
         GrammarVAE    & 0.602 & 0.093 & 0.809 & 0.045 \\
         GraphVAE      & 0.557 & 0.670 & 0.616 & 0.261 \\
         GraphVAE/imp  & 0.562 & 0.520 & 0.758 & 0.179 \\
         GraphVAE NoGM & 0.810 & 0.241 & 0.610 & 0.129 \\
         MolGAN        & \textbf{0.981} & 0.104 & \textbf{0.942} & 0.096 \\
         \hline
         ARAE          & 0.862 & \textbf{0.935} & 0.371 & \textbf{0.299} \\
         \hline
         ARAE (ZINC)   & 0.903 & 1.000 & 1.000 & 0.903  \\
         \hline
         \end{tabular}
         \caption{Performance of benchmark models and our model for the QM9 dataset. Baseline results are taken from De Cao et al.\cite{de2018molgan} 
         We also show the performance of our model for the ZINC dataset in the bottom row. }
         \label{tab:bench_QM9}
    \end{center}
\end{table*}

\begin{figure*}[htb] 
    \includegraphics[width=0.95\textwidth,trim={0cm 0 0cm 0},clip]{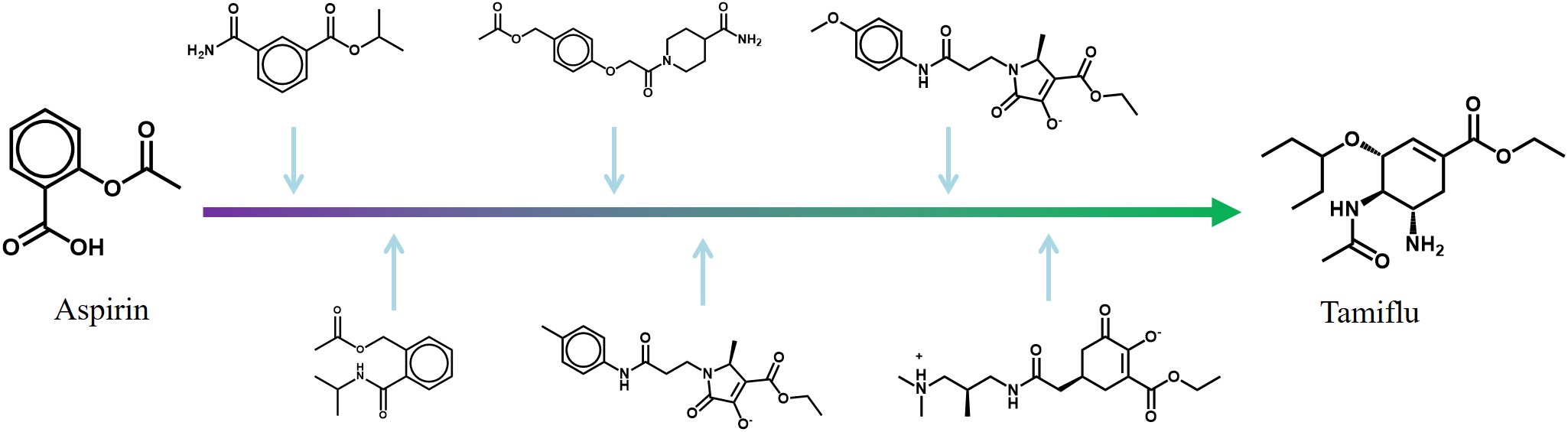}
    \caption{Molecules reconstructed from the latent vectors that appeared in the interpolation among two latent vectors. 
    The starting and ending points are the latent vectors of Aspirin and Tamiflu, respectively.}
    \label{fig:interpolation}
\end{figure*}

Training a generative model based on GANs is often unstable. 
Hence, we first investigated using the QM9 dataset how each evaluation metric changes as the training of our model progresses. 
After each epoch, 10,000 samples were generated, and the four metrics (validity, uniqueness, novelty, and novel/sample) were calculated. 
Figure \ref{fig:metrics} shows the result of the first 80 epochs. 
All evaluation metrics were smoothly converged, indicating that common difficulties in training GANs such as mode collapse or diminished gradient are less problematic in our model. 

We also compare the performance of ARAE to those of ChemicalVAE\cite{gomez2018automatic}, GrammarVAE\cite{kusner2017grammar}, GraphVAE\cite{simonovsky2018graphvae}, and MolGAN\cite{de2018molgan}. 
For an input molecular representation, ChemicalVAE and GrammarVAE used SMILES, while GraphVAE and MolGAN adopted molecular graphs.
Table \ref{tab:bench_QM9} summarizes the validity, uniqueness, novelty, and novel/sample of each model. 
All the models were trained by using the QM9 dataset. 
Overall, ARAE outperformed the others except for novelty. 

As intended, it significantly improved the validity and uniqueness from those of the VAE-based and MolGAN models, respectively, by taking the advantages of both methods. 
The relative low novelty value is due to the low chemical diversity of the QM9 dataset in which molecules are composed of less than ten heavy atoms. 
This limited number of heavy atoms restricts opportunities to generate novel molecules.
On the other hand, ARAE could achieve high novelty for the ZINC dataset (the bottom low of Table 1), because the ZINC dataset spans a huge chemical space. This result further supports the reason of the low novelty of ARAE for QM9.
Though MolGAN also adopted the adversarial training, it showed a high novelty value but by sacrificing the uniqueness, meaning that a high value on one metric can be achieved by sacrificing the others.
Therefore, the ratio of novel molecules among generated samples (novelty/sample) is a more practical metric for performance comparison. Models using graph representation showed relatively higher novel/sample values than models using SMILES. ARAE achieved the highest value of novel/sample in spite of using the SMILES representation. 

One of the major drawbacks of VAE-based generative models is the presence of holes in the latent space due to the approximated posterior, resulting in low validity as noted in Table \ref{tab:bench_QM9}. This problem can often cause the generation of unrealistic molecules at interpolation points between two latent vectors.
In contrast, adversarial training does not impose any analytic form of the posterior. Therefore, it is expected that ARAE can avoid such a problem.
To demonstrate the successful modeling of latent space with the adversarial training, we attempted to generate molecules through an interpolation experiment as shown in Figure \ref{fig:interpolation}.
We obtained 100 latent vectors by linearly interpolating the two seed vectors obtained from Aspirin $\textbf{s}_a$ and Tamiflu $\textbf{s}_b$.
Then, each sampled vector was decoded to generate the corresponding molecule.
All the 100 latent vectors successfully generated valid molecules, and 19 molecules out of them were unique and novel. 
Figure \ref{fig:interpolation} exhibits 6 examples showing smooth change from Aspirin to Tamiflu. 
We also note that high-membered ring molecules which often appear in VAE-based molecular generative models\cite{gomez2018automatic, kusner2017grammar} were not produced by our model.

\subsection{Conditional generation of molecules with CARAE}

\begin{figure*}[htb] 

     \includegraphics[width=0.95\textwidth,trim={0cm 0 0cm 0},clip]{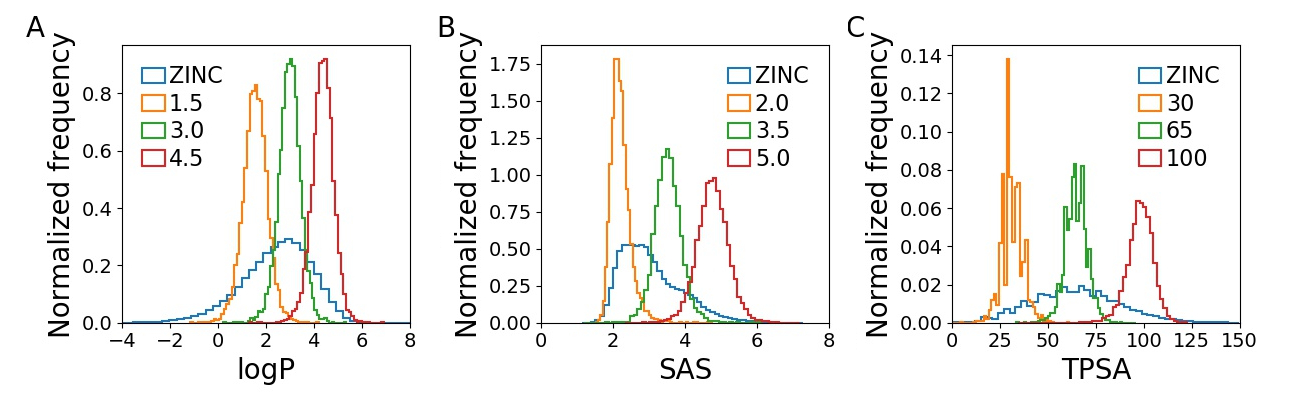}
    \caption{Distributions of molecular property - (a) logP, (b) SAS and (c) TPSA - when molecules are generated by specifying a desired property. Note that the curves labeled with ZINC denote the distribution of each molecular property in the ZINC dataset.}
    \label{fig:property_distributions}
\end{figure*}
In this section, we examine the performance of our model for conditional generation of molecules. 
First, we tested that CARAE can generate molecules with high validity, uniqueness, novelty, and diversity as ARAE does. The performance of CARAE may depend on designated property values, so we compared the performance of CARAE for both random property values and certain designated ones. Three properties (logP, SAS, and TPSA) were controlled simultaneously. Table \ref{tab:bench_ZINC} summarizes the validity, uniqueness, novelty, and diversity of ARAE and CARAE, where both models were trained using the ZINC dataset. CARAE shows comparable performance with that of ARAE except for the high SAS value (5.0). Since the SAS value is related to synthetic accessibility and structural stability, the frequency of valid molecules would be low at a high SAS value. 
The high success rates of the conditional generation are an evidence that the latent space was well separated from multiple target properties, and hence one can readily control molecular structures under given fixed multiple conditions. 

\begin{table*}[htb]
    \begin{center}
        \begin{tabular}{c|cccc}
        \hline
        Method and Condition & Validity & Uniqueness & Novelty & Diversity \\
        \hline
        ARAE                    & 0.903 & 1.000 & 1.000 & 0.909  \\
        \hline  
        CARAE, random$^{\dagger}$          & 0.812 & 1.000 & 1.000 & 0.911  \\
        \hline
        CARAE, (1.5, 2.0, 30)$^{\ddagger}$  & 0.897 & 0.926 & 0.998 & 0.894  \\
        CARAE, (1.5, 2.0, 100) & 0.823 & 0.996 & 1.000 & 0.890  \\
        CARAE, (1.5, 5.0, 30)  & 0.860 & 0.997 & 1.000 & 0.914  \\
        CARAE, (1.5, 5.0, 100) & 0.573 & 1.000 & 1.000 & 0.909  \\
        CARAE, (4.5, 2.0, 30)  & 0.908 & 0.995 & 1.000 & 0.900  \\
        CARAE, (4.5, 2.0, 100) & 0.813 & 1.000 & 1.000 & 0.884  \\
        CARAE, (4.5, 5.0, 30)  & 0.621 & 1.000 & 1.000 & 0.912  \\
        CARAE, (4.5, 5.0, 100) & 0.293 & 1.000 & 1.000 & 0.911  \\
        \hline
        \end{tabular}
    \caption{Performance of ARAE and CARAE trained with ZINC dataset. The diversity of the ZINC testset is 0.915. In this work, we disentangled the information of molecular properties (logP, SAS and TPSA) from latent vectors and incorporated target properties in it later as a condition vector, $\textbf{y}_c$. $^{\dagger}$ For the `CARAE, random', we randomly chose molecular property values and incorporated them in the latent vectors of molecules in the ZINC dataset. $^{\ddagger}$ For the `CARAE, ($\cdot$, $\cdot$, $\cdot$)', we used molecular property values drawn from a Gaussian distribution $N(\cdot, 1.0)$.}
    \label{tab:bench_ZINC}
    \end{center}
\end{table*}
 
\begin{figure}[htb]
    \includegraphics[width=0.48\textwidth,trim={0cm 0 0cm 0},clip]{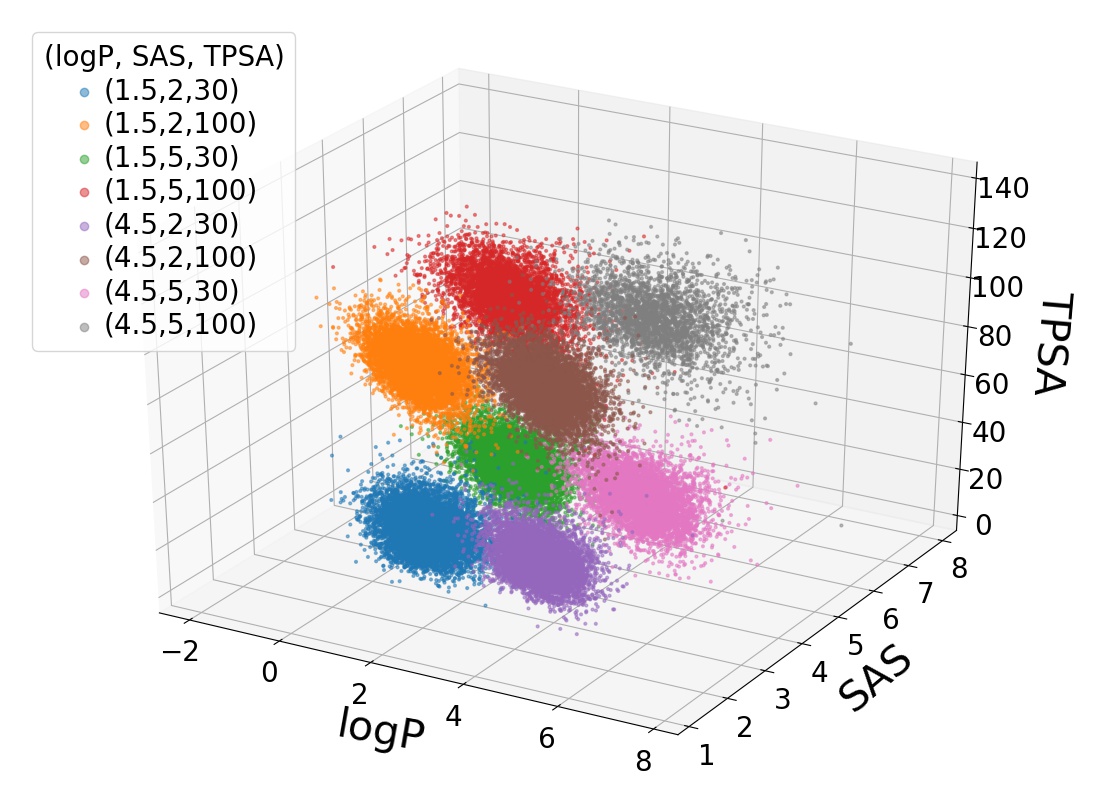}
    \caption{Joint distribution of the logP, SAS and TPSA values of molecules generated with the simultaneous control of the three target properties denoted in the legend.}
    \label{fig:logP_SAS_TPSA}
\end{figure}

We investigated how accurate the target properties of molecules generated by CARAE are.
The conditional generator produced 10,000 molecules with a given target property. 
Figure \ref{fig:property_distributions} compares the normalized frequency of the conditionally-generated molecules for each property and the natural population of the ZINC set molecules; molecular properties for both cases were computed by RDKit.
The logP values of the generated molecules were well localized around the designated values denoted in each panel. The SAS and TPSA values showed relatively broader distributions, but compared to the natural populations, they were also very localized around the given targets. 
Figure \ref{fig:logP_SAS_TPSA} shows the distributions of molecules generated with the simultaneous control of three target properties. Indeed, each distribution was well localized around a given designated point and well separated from each other. These results indicate the high accuracy of our CARAE model for multiple property control as well.

\subsection{\textit{de novo} design of EGFR inhibitors}
Molecular generative models have attracted attention as a promising solution for \textit{de novo} molecular design for new drugs or materials.
To demonstrate the utility of our model, 
we applied it to designing novel inhibitors of an epidermal growth factor receptor (EGFR). 
We obtained active and decoy molecules from the DUD-E database and trained the CARAE model with four target properties: activity against EGFR, logP, TPSA, and SAS. 
Inhibitors were then generated according to the following two scenarios:
\begin{itemize}
    \item Generation with only activity condition: activity = 1. We aimed to generate EGFR-active molecules only. 
    \item Generation with the four conditions: activity = 1, logP = 2.5, SAS = 1.5, and TPSA = 60. 
    It was intended to generate molecules satisfying the Lipinski's rule of five \cite{lipinski20013} and synthesizability.
    
\end{itemize}
\begin{figure*}[htb]
    \includegraphics[width=0.98\textwidth,trim={0cm 0 0cm 0},clip]{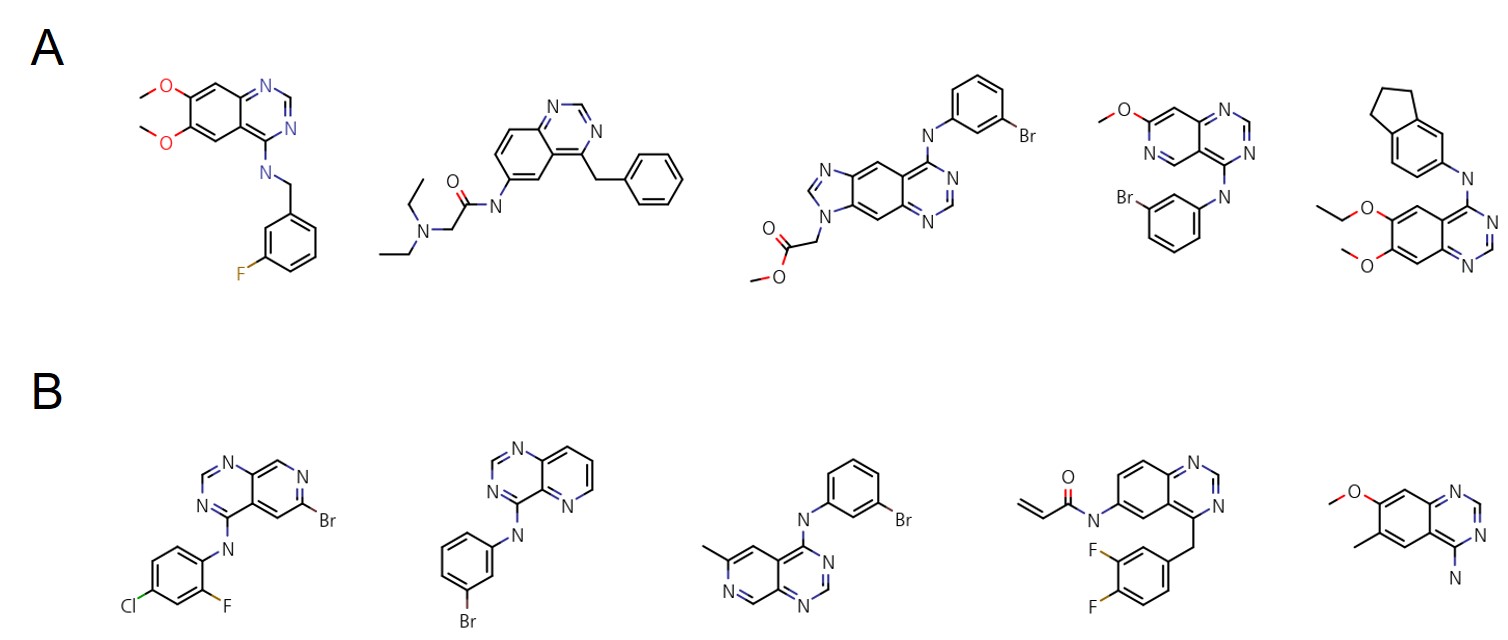}
    \caption{Five generated EGFR inhibitors (A) without additional condition and (B) with additional condition of 'logP = 2.5, SAS = 1.5 and TPSA = 60'. }
    \label{fig:fig6}
\end{figure*}
For both scenarios, we analyzed the scaffolds of the generated valid molecules from 3,500 trials to validate them. 
We used the method proposed by Bemis and Murcko\cite{Bemis1996} to extract the scaffolds of molecules. 
As a result, we could generate new drug candidates with known active scaffolds. 
From the first scenario, 931 new molecules were obtained and 195 molecules out of them shared the scaffolds of known active molecules. 
From the second scenario, 1,067 new molecules were obtained, 88 molecules out of them shared the scaffolds of known active molecules and satisfy the additional conditions. Molecules with 1.5 $<$ LogP $<$ 3.5, 0.5 $<$ SAS $<$ 2.5, and 50 $<$ TPSA $<$ 70 were considered to satisfy the additional conditions. 
The number of successful molecules was decreased in the second scenario because the additional conditions impose more strict constraints than the first scenario. 
Figure \ref{fig:fig6} shows five molecules obtained from each scenario as examples. 

\section{Conclusions}
Conventional molecular generative models are based on either variational autoencoder (VAE) or generative adversarial network (GAN). 
The former often produces invalid molecules due to hole areas in the latent space arisen from an inappropriate approximation of posterior distribution. 
The latter estimates the distribution directly from input data through adversarial training, leading to enhanced validity. 
However, it is problematic in training discrete variables such as molecular structures, resulting in the low uniqueness of generated molecules.  
To address such challenges, we newly propose a molecular generative model built on adversarially regularized autoencoder (ARAE).
It is based on a latent variable model like VAE, but the distribution of latent variables is obtained by adversarial training directly from training data like in GAN instead of approximating with a predefined function. 
Furthermore, it uses continuous latent vectors instead of discrete molecular structures in the adversarial training in order to avoid the difficulty in dealing with discrete variables. 
Based on these facts, it could take the advantages of both VAE and GAN to overcome the limitations of each model.

The benchmark studies in this work provide numerical evidences of the high performance of our ARAE model for molecular generations. First, it outperformed other VAE-based and GAN-based models in terms of the uniqueness and the novel/sample-ratio still with high validity for the QM9 data set.
It also allows smooth interpolation between two molecules in the latent space, which shows the feasibility of the successful modeling of latent space via adversarial training.
Conditional generation of molecules with specific target properties was demonstrated as well. 
Target molecular properties are first disentangled from latent vectors, and desired values are incorporated in the latent vectors to independently control molecular structure and properties. 
Our model showed good performance in conditional generations for both cases of single and multiple properties control.
To demonstrate potential real-world applications, we applied the conditional generation scheme to designing EGFR-inhibitors. 
We could successfully generate new candidate molecules while satisfying drug-like conditions simultaneously. 

Consequently, we believe that ARAE can be a new platform for AI-based molecular design in various chemical applications. 

\section*{Implementation details}

We summarize the configuration of our model in Table \ref{tab:arae_architecture}. 
Each of the encoder and the decoder is composed of a single LSTM layer, and the dimensionality of outputs is 300. 
The LSTM layer of the encoder reads sequential SMILES strings and transforms them to latent vectors. 
For adversarial training, we use two fully-connected layers with a hidden dimension of 300 for ZINC (200 for QM9) for the generator and the discriminator networks.
The predictor network is also composed of two fully-connected layers with a hidden dimension 300 for ZINC (200 for QM9). 
We uploaded our code on github \url{https://github.com/gicsaw/ARAE_SMILES}.

\begin{table*}[htb]
    \begin{center}
    \resizebox{\textwidth}{!}{
         \begin{tabular}{ccccc}
         \hline
                        & Layer configuration            & Nonlinearity & Dimension (QM9) & Dimension (ZINC) \\
         \hline
         Encoder        & One LSTM layer                 & -            & 200             & 300 \\
         Decoder        & One LSTM layer + Softmax layer & -            & 200             & 300 \\
         Generator      & Two fully-connected layers     & ReLU         & 200             & 300 \\
         Discriminator  & Two fully-connected layers     & LReLU        & 200             & 300 \\
         Predictor      & Two fully-connected layers     & ReLU         & -               & 300 \\
         \hline
         \end{tabular}
         \caption{Configuration of the model used in this work. Note that the predictor is only used for conditional generations with CARAE.  }
         \label{tab:arae_architecture}
    }
    \end{center}
\end{table*}

Table \ref{tab:arae_optimizer} describes the hyperparameters and the optimizers of ARAE.

\begin{table*}[htb]
    \begin{center}
         \begin{tabular}{cccc}
         \hline
          & Variable parameters & Optimizer& Learning rate \\
         \hline
         Auto-Encoder   & Encoder and Decoder & SGD & 1.0 \\
         Generator   & Generator & Adam & $1.0 \times 10^{-5}$ \\
         Critic   & Critic & Adam & $2.0 \times 10^{-6}$ \\
         Predictor   & Predictor & Adam & 1.0 \\
         Disentanglement   & Encoder & Adam & 1.0 \\
         \hline
         \end{tabular}
         \caption{Optimizers and prarameters of ARAE. }
         \label{tab:arae_optimizer}
    \end{center}
\end{table*}

\section*{Acknowledgements}
This work was supported by the National Research Foundation of Korea (NRF) grant funded by the Korea government (MSIT)(NRF-2017R1E1A1A01078109).

\section*{Author contributions}
J.L., S.R. and W.Y.K. conceived the idea, J.L. and S.H.H. did the implementation and run the simulation. 
All the authors analyzed the results and wrote the manuscript together.

\section*{Conflicts of interest}
The authors declare no competing financial interests.

\bibliography{achemso-demo}

\end{document}